\documentclass[traditabstract]{aa}

\usepackage{natbib}
\bibpunct{(}{)}{;}{a}{}{,} 
\usepackage{amssymb}
\usepackage{url}
\usepackage{graphicx}
\usepackage{booktabs}           
\usepackage[point,norounding]{rccol}

\newcommand{\msun}{\mbox{$M_\odot$}}
\newcommand{\lsun}{\mbox{$L_\odot$}}

\newcommand{\GiveRef}[1]{\citetalias{#1}: \citet{#1}}
\newcommand{\salpeter}{Salpeter}
\newcommand{\chabrier}{Chabrier}

\newcommand{\mstar}{M_{\rm star}}

\begin{document}

 \title{Dust production $680$--$850$ million years after the Big Bang}
 
\titlerunning{Dust production $680$--$850$ million years after the Big Bang}
\authorrunning{Micha{\l}owski}

\author{Micha{\l}~J.~Micha{\l}owski
        }

\institute{
SUPA\thanks{Scottish Universities Physics Alliance}, Institute for Astronomy, University of Edinburgh, Royal Observatory, Blackford Hill, Edinburgh, EH9 3HJ, UK, {\tt mm@roe.ac.uk} 
\label{inst:roe}
}

   \date{Received 12 January 2015; accepted 23 March 2015}

\abstract{%
Dust plays an important role in our understanding of the Universe, 
but it is not obvious yet how  the dust in the distant universe was formed.
I derived the dust yields per asymptotic giant branch (AGB) star and per supernova (SN) required to explain dust masses of galaxies at $z=6.3$--$7.5$ ($680$--$850$ million years after the Big Bang) for which dust emission has been detected (HFLS3 at $z=6.34$, ULAS J1120+0641 at $z=7.085$, and A1689-zD1 at $z=7.5$),
 or unsuccessfully searched for. I found very high required yields, implying that AGB stars could not contribute substantially to dust production at these redshifts, and that SNe could explain these dust masses, but only if they do not destroy most of the dust they form (which is unlikely given the upper limits on the SN dust yields derived for  galaxies where dust is not detected). This suggests that the grain growth in the interstellar medium is likely required at these early epochs.
}


\maketitle

\section{Introduction}
\label{sec:intro}

\defcitealias{riechers13}{1}
\defcitealias{mortlock11}{2}
\defcitealias{venemans12}{3}
\defcitealias{watson15}{4}
\defcitealias{hu02}{5}
\defcitealias{schaerer14}{6}
\defcitealias{ouchi09}{7}
\defcitealias{hirashita14}{8}
\defcitealias{ouchi13}{9}
\defcitealias{bradley12}{10}
\defcitealias{iye06}{11}
\defcitealias{finkelstein13}{12}
\begin{table*}
\caption{Properties of the galaxies in my sample. \label{tab:sample}   }
\begin{scriptsize}
\begin{tabular}{lcccccccccccl}
\hline\hline
Name & z & \multicolumn{3}{c}{$\log (M_{\rm dust} / \msun)$}  & \multicolumn{3}{c}{$\log (M_{\rm star} / \msun)$} & $\log (M_{\rm dyn} / \msun)$ & $\log (M_{gas} / \msun)$ & Ref\\
\cmidrule(lr){3-5} \cmidrule(lr){6-8}
\hline
HFLS3 & 6.340 & $8.75^{+0.09}_{-0.11}$ & $\cdots$ & $\cdots$ &$10.57^{+0.30}_{-0.30}$ &$10.04^{+0.10}_{-0.13}$ &$11.46^{+0.30}_{-0.30}$\tablefootmark{a} &$11.43$ &$11.02^{+0.04}_{-0.04}$\phantom{\tablefootmark{b}} &\citetalias{riechers13} \\
ULASJ1120+0641 & 7.085 & $7.72^{+0.09}_{-0.11}$ &$7.87^{+0.09}_{-0.11}$ &$8.41^{+0.10}_{-0.13}$ & $\cdots$ & $\cdots$ & $\cdots$ &$11.00$ &$10.08^{+0.10}_{-0.13}$\tablefootmark{b} &\citetalias{mortlock11},\citetalias{venemans12} \\
A1689-zD1 & 7.500 & $7.60^{+0.30}_{-0.30}$ &$7.48^{+0.22}_{-0.48}$ &$7.85^{+0.11}_{-0.15}$ &$9.23^{+0.05}_{-0.06}$ &$9.30^{+0.07}_{-0.09}$ &$9.40^{+0.07}_{-0.09}$ & $\cdots$ & $\cdots$ &\citetalias{watson15} \\
\hline
HCM6A & 6.560 & $<7.44$ & $\cdots$ & $\cdots$ &$9.24^{+0.30}_{-0.30}$ & $\cdots$ & $\cdots$ & $\cdots$ & $\cdots$ &\citetalias{hu02},\citetalias{schaerer14} \\
Himiko & 6.595 & $<6.94$ & $\cdots$ & $\cdots$ &$10.29^{+0.06}_{-0.06}$ &$9.92^{+0.02}_{-0.02}$ & $\cdots$ & $\cdots$ & $\cdots$ &\citetalias{ouchi09},\citetalias{hirashita14},\citetalias{ouchi13} \\
A1703-zD1 & 6.800 & $<7.19$ & $\cdots$ & $\cdots$ &$8.94^{+0.30}_{-0.30}$ & $\cdots$ & $\cdots$ & $\cdots$ & $\cdots$ &\citetalias{bradley12},\citetalias{schaerer14} \\
IOK-1 & 6.960 & $<7.27$ & $\cdots$ & $\cdots$ &$9.44^{+0.30}_{-0.30}$ & $\cdots$ & $\cdots$ & $\cdots$ & $\cdots$ &\citetalias{iye06},\citetalias{schaerer14} \\
z8-GND-5296 & 7.508 & $<8.11$ & $\cdots$ & $\cdots$ &$9.44^{+0.30}_{-0.30}$ & $\cdots$ & $\cdots$ & $\cdots$ & $\cdots$ &\citetalias{finkelstein13},\citetalias{schaerer14} \\
\hline
\end{tabular}
\end{scriptsize}
\tablefoot{ 
Unless noted otherwise the values are from the references listed in the last column. Stellar masses are listed for the \citet{chabrier03} IMF.
\tablefoottext{a}{I calculated the stellar mass from the photometry \citep{riechers13} using the {\sc Grasil} code.}
\tablefoottext{b}{I calculated the gas mass from the $L_{\rm CII}$ luminosity \citep{venemans12} using the empirical calibration of \citet{swinbank12}: $M_{\rm gas}=(10\pm2)(L_{\rm CII}/\lsun)$.}
}
References:  \GiveRef{riechers13},  \GiveRef{mortlock11}, 
\GiveRef{venemans12},  \GiveRef{watson15},  \GiveRef{hu02}, 
\GiveRef{schaerer14},  \GiveRef{ouchi09},  \GiveRef{hirashita14}, 
\GiveRef{ouchi13},  \GiveRef{bradley12},  \GiveRef{iye06}, 
\GiveRef{finkelstein13}, 
\end{table*}

Dust plays an important role in our understanding of the Universe, both hiding the regions of intense star formation and allowing this information to be regained by the observations of its thermal emission. It is, however, not obvious yet how  the dust in the distant universe was formed, as dust formation requires specific conditions of relatively low temperature and high density. These conditions are met in atmospheres of asymptotic giant branch (AGB) stars and expelled shells of supernova (SN) remnants (see \citealt{gall11c}, for a review). Another possibility is that these stellar sources produce only dust seeds, and most  dust mass accumulation happens in the interstellar medium \citep[ISM;][]{draine79}.

Dust has been observed to form in numerous AGB stars \citep{meixner06,matsuura09,matsuura13,sloan09,srinivasan09,boyer11,boyer12,riebel12}, and 
it was shown theoretically that one AGB star can produce up to $\sim0.04\,\msun$ of dust \citep{morgan03,ferrarotti06,ventura12,nanni13,nanni14,schneider14}. 
 This maximum dust mass is returned only for a narrow range  of the mass of an AGB star progenitor $(\sim4\,\msun)$ and for super-solar metallicity \citep[Table~3 in][]{ferrarotti06}. For solar and subsolar metallicities, up to $\sim0.02\,\msun$ of dust can be produced by an AGB star, with a typical value of $\sim0.001\,\msun$  across a wider mass range.
It has been claimed that AGB stars
 produce most of the stellar dust (as opposed to dust grown in the ISM)
in the Milky Way and local galaxies \citep{gehrz89,zhukovska13}, but also in high-redshift quasars \citep{valiante09,valiante11}.

Theoretical models predict that a SN can produce at most $\sim1.3\,\msun$ of dust \citep{todini01,nozawa03}, but likely only $\lesssim0.1\,\msun$ survives the associated shocks  and is released  into the ISM \citep{bianchi07,cherchneff10,gall11,lakicevic15}. 
Large  amounts of dust have been found in the remnants of  \object{Cassiopeia~A} \citep{dunne03,dunne09casA} and Kepler \citep{morgan03b,gomez09} ($\sim1\,\msun$);  SN 1987A  \citep[$\sim0.4$--$0.7\,\msun$;][]{matsuura11,indebetouw14}, and the Crab Nebula \citep[$\sim0.02$--$0.2\,\msun$][]{gomez12,temim13}. 
 However, the high dust yields for Cassiopeia A and Kepler are still controversial \citep{dwek04,krause04,gomez05,wilson05, blair07,sibthorpe09,barlow10}, and dust masses for other SN remnants are typically much lower, of the order of
$10^{-3}$--$10^{-2}\,\msun$
 \citep{green04,borkowski06,sugerman06,ercolano07,meikle07,rho08,rho09,kotak09,lee09,sakon09,sandstrom09,wesson09,gallagher12,temim12}. 
 Some of this discrepancy can be attributed to different observed wavelengths used by different authors. If the dust is relatively cold, then mid-infrared {\it Spitzer} observations do not probe the dominant cold dust component, and would underestimate the total dust mass. Moreover,
this difference may be connected with the evolutionary stage of a SN remnant.
 Recently, \citet{gall14} found very large grains in the circumstellar medium of a SN just a few months after the explosion, and claimed that such grains are able to survive the shocks and help to condense larger amounts of dust a few tens of years later.
 
It has been claimed that SNe are the main dust producers at high redshifts, when at least some of the intermediate-mass stars had not had enough time to reach the AGB phase \citep{dwek07, dwek11,dwek11b,dwek14,michalowski10smg4,michalowski10qso,gall11b,hjorth14,rowlands14b}.
However, using the ALMA non-detection of a $z\sim6.6$ Ly$\alpha$ emitter 
\citep{ouchi13}, \citet{hirashita14} found that only $<0.15$--$0.45\,\msun$ of dust can be returned to the ISM by a SN.

Because of this difficulty in finding an efficient dust production mechanism by stellar sources, the grain growth in the ISM has been claimed to be necessary for 
the Milky Way \citep{draine79,dwek80,draine90,draine09}, local galaxies \citep{matsuura09,grootes13}, and distant galaxies  \citep{dwek07,michalowski10smg4,michalowski10qso,hirashita11,mattsson11,mattsson12,kuo12,asano13,calura14,hjorth14,rowlands14b,valiante14,nozawa15}. The grain growth timescale  is relatively short  \citep[a few tens of Myr;][]{draine90,draine09,hirashita00,hirashita12,zhukovska08}, which is consistent with very little variation in the metal-to-dust ratios in gamma-ray burst (GRB) and quasar absorbers, as this implies very rapid dust production soon after metal enrichment \citep{zafar13}.
 However,  metal-to-dust ratios are higher at metallicities well below 0.1 solar, as shown by IZw18 (Fig.~2 in \citealt{zafar13}) and by a steeper slope of the gas-to-dust ratio for low metallicities (Fig.~4 in \citealt{remyruyer14}). This is connected with the grain growth timescale being much longer in the metal-poor environments \citep[Eq.~(20) of][]{asano13} when stellar sources are the main dust producers.

Apart from using the total dust budget  arguments, these three possibilities are very difficult to distinguish because doing so would require precise measurements of dust properties (grain size distribution and chemical composition). Therefore, SN-origin dust has only been claimed based on  extinction curves of a $z\sim6.2$ quasar (\citealt{maiolino04,gallerani10}; but see \citealt{hjorth13}) and two GRB host galaxies at $z\sim6.3$ (\citealt{stratta07b}; but see \citealt{zafar10,zafar11}) and  $z\sim5$ \citep{perley10}.

The objective of this paper is to investigate whether SNe and AGB stars are efficient enough to form dust at $6.3<z<7.5$ ($680$--$850$ million years after the Big Bang), or if  grain growth in the ISM is required.
I use a cosmological model with $H_0=70$ km s$^{-1}$ Mpc$^{-1}$, $\Omega_\Lambda=0.7$, and $\Omega_m=0.3$.

\section{Sample}
\label{sec:sample}

\begin{figure*}
\begin{center}
\includegraphics[width=\textwidth]{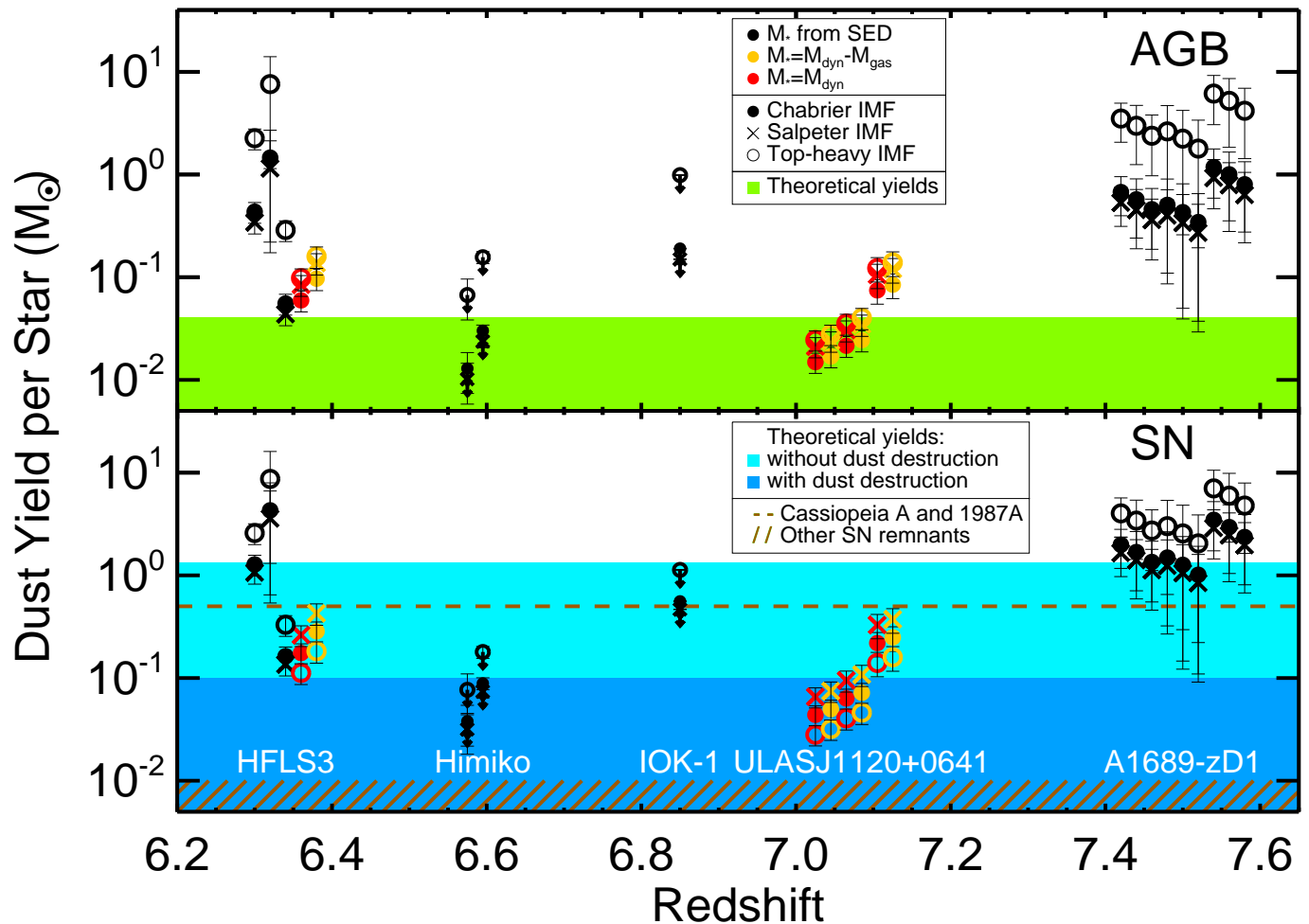}
\end{center}
\caption{Dust yield per AGB star ({\it top}) and per SN ({\it bottom}) required to explain dust masses measured in $z>6.3$ galaxies for all possible combinations of dust and stellar masses (Table~\ref{tab:sample}; the points were shifted slightly in redshift for clarity). Symbols are colour-coded according to the way the stellar mass was estimated: from spectral energy distribution modelling ({\it black}), assuming that the stellar mass is the difference between the dynamical and gas masses ({\it yellow}), or that it is equal to the dynamical mass ({\it red}). {\it Filled circles}, {\it crosses}, and {\it open circles} denote the assumption of the Chabrier, Salpeter ($\alpha=2.35$), and top-heavy ($\alpha=1.5$) initial mass functions, respectively. 
 Small symbols with arrows correspond to galaxies with only upper limits for dust masses (lower part of Table~\ref{tab:sample}).
}
\label{fig:yield}
\end{figure*}

For this study I selected all galaxies at $z>6.3$ for which dust emission has been detected (with the exception of \object{SDSS J1148+5251}, a $z=6.42$ quasar that was a subject of the study in \citealt{michalowski10qso}, and MACS1149-JD, a $z\sim9.6$ Lyman break galaxy [LBG], because its association with the 2\,mm GISMO object has not yet been confirmed; \citealt{dwek14}). The sample includes \object{HFLS3}, a  red {\it Herschel}-selected $z=6.34$ galaxy \citep{riechers13}; \object{ULAS J1120+0641}, a colour-selected $z=7.085$ quasar \citep{mortlock11,venemans12}; and \object{A1689-zD1}, a lensed $z=7.5$ Lyman break galaxy \citep{watson15}. 

 Additionally, I selected galaxies at $z>6.3$ for which the dust emission has been unsuccessfully searched for:
$z = 6.56$ lensed  Ly$\alpha$ emitter \object{HCM6A} \citep[][magnification factor of $4.5$]{hu02,kanekar13},
$z=6.8$ lensed LBG \object{A1703-zD1} \citep[][magnification factor of $9$]{bradley12,schaerer14},
$z=6.595$ Ly$\alpha$ emitter \object{Himiko} \citep{ouchi09,ouchi13},
$z=6.96$  Ly$\alpha$ emitter \object{IOK-1} \citep{iye06,ota14}, and
$z=7.508$ LBG \object{z8-GND-5296} \citep{finkelstein13,schaerer14}.

Table~\ref{tab:sample} presents the relevant properties of these galaxies, which are adopted from the references listed in the last column, except for the third stellar mass estimate for HFLS3 and the gas mass estimate for ULASJ1120+0641.
Dust masses from \citet{schaerer14} are adopted with the $T_{\rm dust}=35$\,K option.
 Multiple values for a given property denote different assumptions adopted when deriving it (temperatures and emissivity index for dust masses, and star formation histories for stellar masses).
I converted all dust mass estimates to a common mass absorption coefficient  of $\kappa_{1200\,\mu{\rm m}}=0.67\,{\rm cm}^2 {\rm g}^{-1}$, and stellar masses to the {\chabrier} IMF. 

For the stellar mass estimation for HFLS3 I used the photometry presented in \citet{riechers13} and applied the SED fitting method detailed in \citet[][see therein the discussion of the derivation of galaxy properties and typical uncertainties]{michalowski08,michalowski09,michalowski10smg,michalowski10smg4}, which is based on 35\,000 templates from the library of \citet{iglesias07} and some templates from \citet{silva98} and \citet{michalowski08}, all of which were developed using {\sc Grasil}\footnote{\url{http://adlibitum.oats.inaf.it/silva/grasil/grasil.html}} \citep{silva98}. They are based on numerical calculations of radiative transfer within a galaxy, which is assumed to be a triaxial axisymmetric system with diffuse dust and dense molecular clouds, in which stars are born.

I estimated the gas mass of ULASJ1120+0641 from the [\ion{C}{II}] luminosity, $L_{\rm CII}$ \citep{venemans12}, using the empirical calibration of \citet{swinbank12}: $M_{\rm gas}=(10\pm2)(L_{\rm CII}/\lsun)$. The uncertainty of this estimate has little effect on my result, as this mass corresponds only to $\sim10\%$ of the dynamical mass estimate.

\section{Method}
\label{sec:method}

The redshifts of the galaxies considered here correspond to at most $\sim500\,$Myr of evolution, depending on the formation redshift. Hence, for AGB stars I only considered stars more massive than $3\,\msun$, as less-massive stars have not had time to leave the main sequence and to produce dust \citep[the main-sequence lifetime is $10^{10}\,\mbox{yr} \times [M/\msun\mbox{$]$} ^{-2.5}$; e.g.,][]{kippenhahn90}.

I calculated the dust yields per  AGB star and per SN (amount of dust formed in ejecta of one star) required to explain the inferred dust masses in these galaxies as described in \citet{michalowski10smg4,michalowski10qso}. Namely, the number of stars with masses between $M_0$ and $M_1$ in the stellar population with a total stellar mass of $\mstar$ was calculated as $N(M_0$--$M_1)=\mstar \int_{M_0}^{M_1} M^{-\alpha} dM / \int_{M_{\rm min}}^{M_{\rm max}} M^{-\alpha}  M dM$, where $(M_0,M_1)=(3,8)\,\msun$ for AGB stars and $(8,40)\,\msun$ for SNe, respectively.  Depending on the galaxy, $\mstar$ was estimated in three ways: derived from the SED modelling  (reported in Cols. 6--8 of Table~\ref{tab:sample}), assumed to be equal to $M_{\rm dyn}-M_{\rm gas}$ and equal to $M_{\rm dyn}$. The average dust yield per  star is $M_{\rm dust} /  N(M_0$--$M_1)$. With these assumptions this yield  is $f\times(M_{\rm dust}/\mstar)$, where $f=\int_{M_{\rm min}}^{M_{\rm max}} M^{-\alpha}  M dM / \int_{M_0}^{M_1} M^{-\alpha} dM$ (Table~\ref{tab:imf}).

I adopted an initial mass function (IMF) with $M_{\rm min}=0.15$, $M_{\rm max}=120\,\msun$, and three possible shapes: \citet{chabrier03}, \citet{salpeter} with a slope $\alpha=2.35$, and  top heavy with a slope $\alpha=1.5$.

\section{Results}
\label{sec:res}

Figure~\ref{fig:yield} presents the dust yields per AGB star and per SN required to explain the observed dust masses using all possible combinations of stellar and dust masses presented in Table~\ref{tab:sample} and three choices of IMFs. 

 The maximum theoretical yield for an AGB star is shown as $0.04\,\msun$ because, as explained above, it is a strict upper limit. For SNe I plot the theoretical value of a dust yield with no dust destruction ($1.3\,\msun$), and the expected dust mass returned to the ISM if dust destruction is taken into account ($0.1\,\msun$).

 Among the three galaxies with dust detections Fig.~\ref{fig:yield} shows that only the dust mass of ULAS J1120+0641 can be explained by both AGB stars (required yields $0.015$--$0.04\,\msun$ for the low dust mass options), or SNe with dust destruction taken into account (required yields $0.03$--$0.1\,\msun$), and only if the real dust yields per star are close to the maximum theoretically allowed values. However, AGB stars are not able to produce dust in two other galaxies (required yields $0.1$-$1\,\msun$), and SNe would need to return most of the formed dust without destroying it (required yields $0.2$--$1\,\msun$). 

Most of the galaxies not detected at the dust continuum (lower part of Table~\ref{tab:sample}) have  stellar masses that are too small to provide meaningful constraints on AGB and SN dust yields, in the sense that the dust mass expected from the resulting number of these stars is much lower than the measured dust mass upper limits.
 In Fig.~\ref{fig:yield}, I show only the upper limits for Himiko and IOK-1, which are massive enough, so that the expected number of AGB stars and SNe is high enough to give reasonable constraints on the maximum dust yield per star.   I found comparable results for Himiko with \citet{hirashita14}; in this galaxy each SN could have produced only $<0.2\,\msun$ of dust assuming the top-heavy IMF, and $<0.03$--$0.09\,\msun$ for the {\chabrier} and {\salpeter} IMFs. In IOK-1, one SN could have produced $<1.1\,\msun$ of dust with the top-heavy IMF, and $<0.5\,\msun$ with the {\chabrier} and {\salpeter} IMFs. The dust yield upper limits for AGB stars are not constraining (above the theoretical limits), except for Himiko with the {\chabrier} and {\salpeter} IMFs, for which I derived $<0.01$--$0.03\,\msun$ of dust per AGB star.

Table~\ref{tab:imf} presents the factor $f$ by which the $M_{\rm dust}/\mstar$ ratio needs to be multiplied for a given IMF to obtain the dust yield per star, and the factor by which the SED-derived stellar masses were multiplied for this IMF (the stellar mass factor for the top-heavy IMF with $\alpha=1.5$ is from \citealt{sternberg98}). These factors can be quickly used to calculate the average dust yields, and they only depend on the adopted IMF and stellar mass ranges of dust producers.

\section{Discussion}
\label{sec:discussion}

\begin{table}
\caption{Initial mass functions assumed. \label{tab:imf}   }
\begin{tabular}{lcccc}
\hline\hline
IMF & $f_{\rm AGB}$\tablefootmark{a}  & $f_{\rm SN}$\tablefootmark{a}  & $\mstar / M_{\rm Chab}$\tablefootmark{b} & sym.\tablefootmark{c}\\
\hline
Chabrier & 29 & \phantom{1}84 & 1\phantom{.00} & $\bullet$\\
Salpeter &  41 & 127 & 1.8\phantom{0}  & $\times$\\
Top-heavy & 47 & \phantom{1}54 & 0.32 & $\circ$  \\
\hline
\end{tabular}
\tablefoot{ 
\tablefoottext{a}{IMF-dependent factor, defined such that the required dust yield per star is $f\times(M_{\rm dust}/\mstar)$. The assumed stellar mass ranges are $3$--$8\,\msun$ for AGB stars and $8$--$40\,\msun$ for SNe.}
\tablefoottext{b}{Ratio of the stellar mass derived using this IMF to the {\chabrier} IMF.}
\tablefoottext{c}{Symbol used in Fig.~\ref{fig:yield}.}
}
\end{table}

The dust yields per star shown in Fig.~\ref{fig:yield} imply that dust at $z\sim6.3$--$7.5$ could not have been  entirely formed by AGB stars, and that SNe are efficient enough only if they do not destroy the pre-existing dust and most of the dust they produce. 
 However, such high SN dust yields are  difficult to reconcile with upper limits derived for galaxies not detected at the dust continuum (Himiko and IOK-1, see Fig.~\ref{fig:yield} and \citealt{hirashita14}).
 This suggests that most of the dust mass in these galaxies had been formed in the ISM. This is in line with similar claims for galaxies at lower redshifts. If the grain growth in the ISM is as fast as $\sim10\,$Myr, then even at these redshifts there is a sufficient amount of time for such dust mass accumulation 
 (metallicities of the galaxies considered here are likely to be above the critical value  of $\sim0.1$--$0.3$ solar, the point at which  the contribution of the grain growth falls below that of  stellar sources because of the long timescale of this growth; \citealt{asano13,remyruyer14}).

These conclusions obviously depend on how accurately the relevant masses were measured. I show here that they are robust because all my critical assumptions are conservative in the sense that they tend to bias the resulting required dust yield per star towards low values:
{\it i)} I assumed a relatively high value of the mass absorption coefficient  $\kappa$ \citep[compare with][]{alton04}, higher than what is usually assumed for distant galaxies, resulting in a systematically low $M_{\rm dust}$; 
{\it ii)} I included the stellar mass alternatives, which represent virtually the maximum value, and I assumed that the entire dynamical mass is composed of stars (red symbols in Fig.~\ref{fig:yield}) and assumed a double-component star formation history (SFH) in the SED modelling, which has been shown to result in higher stellar masses than single component SFHs \citep{michalowski12mass,michalowski14mass};
{\it iii)} by including {\it all} stars more massive than $3\,\msun$ (with the main-sequence lifetime of $\lesssim650\,$Myr),  I implicitly assumed that the entire stellar mass was formed at the very beginning of the galaxy evolution, because otherwise a fraction of stars formed later had not finished their main-sequence stage and so should not be taken into account  (the final dust mass depends on the SFHs, not only on the final stellar mass). This extreme  SFH is not ruled out by the data, but is highly unlikely;
{\it iv)} the theoretically  allowed (coloured) regions in Fig.~\ref{fig:yield} represent the maximum dust yield per star, which does not apply to all stars in a given mass range, so the average dust yield per star is  lower;
and {\it v)} I implicitly assumed that dust injected into the ISM is not subsequently destroyed by astration, interstellar shocks, or outflows,  otherwise the produced dust masses would be higher than the currently existing dust masses and, in turn, the required stellar dust yields would be higher.
In each of these cases different (more realistic) assumptions than those chosen here would result in even higher required dust yields, making my conclusions even stronger.

  Moreover, in this analysis I considered AGB stars and SNe separately as dust producers. In reality, both sources operate simultaneously, at least when AGB star progenitors evolve off the main sequence. If the contribution of AGB stars and SNe is equal, then this would bring the points in Fig.~\ref{fig:yield} down by a factor of two, which would not change my conclusions.

The choice of the IMF does not influence these conclusions strongly. For the same dust and stellar masses the {\salpeter} IMF results in yields that are a factor of $\sim1.5$ higher  than for the {\chabrier} IMF (see factor $f$ in Table~\ref{tab:imf}), which makes it even more difficult to explain the observed dust masses. Since the  SED-derived stellar masses are a factor of $\sim1.8$ higher for the {\salpeter} IMF, the resulting yields using these SED-derived masses  are almost identical to those derived with the {\chabrier} IMF. 

Adopting the top-heavy IMF (and keeping masses the same) would increase the required dust yield per AGB star by a factor of $\sim1.6$ ($f_{\rm AGB}$ in Table~\ref{tab:imf}), virtually ruling out their contribution to the dust production in these galaxies. On the other hand, in the top-heavy IMF the fraction of the most massive stars is higher, so the required dust yields per SN are lower by a factor of $\sim0.65$ ($f_{\rm SN}$ in Table~\ref{tab:imf}). However, the SED-derived stellar masses are $3$ times lower, so all in all the number of these massive stars is lower than in the {\chabrier} IMF, so the required dust yields are higher. Hence,  the {\chabrier} IMF is the most conservative assumption (resulting in the lowest required yields).

 Finally, it needs to be pointed out that the three galaxies with detected dust emission are very different to each other, as the sample includes a dusty starburst, a quasar, and a more typical star-forming galaxy (the galaxies with dust non-detections also fall into this last category). This means that their dust properties and origins may be vastly different, even at similar stellar masses. Hence, the relative contribution of AGB stars, SNe, and the ISM grain growth may be different in these sources.

\section{Conclusions}
\label{sec:conclusion}

Using all three galaxies at $z=6.3$--$7.5$ ($680$--$850$ million years after the Big Bang) for which dust emission has been detected (HFLS3 at $z=6.34$, ULAS J1120+0641 at $z=7.085$, and A1689-zD1 at $z=7.5$)
 and galaxies at the same redshift interval for which dust emission has not been detected, I found that AGB stars could not contribute substantially to the dust production at these redshifts, and that SNe could explain their dust masses, but only if they do not destroy most of the dust they form . This last constraint  is unlikely given the upper limits on the SN dust yields derived for  galaxies where dust has not been detected. This suggests that  grain growth in the interstellar medium is likely to be required at these early epochs of the evolution of the Universe.


\begin{acknowledgements}

I thank Joanna Baradziej, Jens Hjorth, Darach Watson, and my anonymous referee 
for useful suggestions,
and acknowledge the support of the UK Science and Technology Facilities Council.
This research has made use of  
the NASA's Astrophysics Data System Bibliographic Services;
and the  Edward Wright Cosmology Calculator \url{www.astro.ucla.edu/~wright/CosmoCalc.html} \citep{wrightcalc}.

\end{acknowledgements}



\end{document}